\documentclass[12pt]{iopart}
\input xy
\xyoption{all}
\usepackage{graphicx}
\usepackage{color}

\newcommand{\U}{\mathcal{U}}

\renewcommand{\1}{{\mathbf 1}}

\renewcommand{\L}{\mathcal{L}}
\renewcommand{\S}{\mathcal{S}}
\renewcommand{\SS}{\mathbf{S}}

\begin{document}

\title[Geometric uncertainty relation for quantum ensembles]{Geometric uncertainty relation for quantum ensembles}
\author{Hoshang Heydari and Ole Andersson }

\address{Department of Physics, Stockholm  University, SE-106 91 Stockholm
, Sweden}
\ead{hoshang@fysik.su.se}
\begin{abstract}
Geometrical structures of quantum mechanics provide us with new insightful results about the
nature of quantum theory. In this work we consider mixed quantum states represented by
finite rank density operators. We review our geometrical framework that provide the
space of density operators with Riemannian and symplectic structures, and we derive a geometric
uncertainty relation for observables acting on mixed quantum states. We also give an
example that visualizes the geometric uncertainty relation for spin-$\frac{1}{2}$ particles.
\end{abstract}

\section{Introduction}
The phase spaces of classical and quantum mechanical systems are symplectic manifolds, and in both cases observables give rise to symplectic flows \cite{Kibble_1979,Ashtekar_etal1998,Brody_etal1999}.
However, quantum systems exhibit characteristics that have no classical counterparts.
One is the impossibility to fully predict results of measurements.
In classical mechanics, the results of the measurements are completely predictable.
But in quantum mechanics  the actual value of an observable cannot be known prior to measurement, and there is a lower bound to the precision with which values of pairs of observables can be known simultaneously which is called uncertainty principles or relations.   Pioneering works on the uncertainty relation includes \cite{Heisenberg1927, Kennard1927, Weyl1928,Robertson_1929, Schrodinger_1930}. Recently several other versions of  uncertainty relation have been considered in \cite{Lou2003,Lou2005, Park2005,Dodonov2002,Wehner_etal2010,Ingarden1973,Dodonov_etal1980,Tarasov2013}.
The uncertainty relation not only is one of the most importance  and central topic in foundations of quantum mechanics but also it has  many applications in quantum information \cite{Jaeger,Dariano, Khrennikov,Sen, Majumdar}. In particular, Robertson-Schr\"{o}dinger uncertainty relation \cite{Robertson_1929, Schrodinger_1930} has been used for discrimination between entangled  and separable states \cite{Nha} and in the domain of discrete variable to distinguish pure states from mixed states \cite{Mal}.

The space of a pure quantum state is projective Hilbert space equipped with the Fubini-Study metric. The real and imaginary parts of the Fubini-Study  metric  equips the projective Hilbert space with Riemannian and symplectic structures.
Ashtekar and Schilling \cite{Ashtekar_etal1998} have shown that for observables acting on a system in a pure state, the Robertson-Schr\"{o}dinger uncertainty relation \cite{Robertson_1929,Schrodinger_1930} can be
expressed entirely in terms of the Riemann and Poisson brackets of the observable's  expectation value functions.

Recently, we have introduced a geometric framework for density operators which have resulted in many interesting topics such as geometric phases, uncertainty relations, quantum speed limits, distance measure, and a characterization of optimal Hamiltonians \cite{Ole1,Ole2,Ole3,Ole4,Ole5,Ole6}.

In this  paper we discuss an uncertainty relation for mixed quantum states based on geometrical structures of the space of density operators which is a generalization of Ashtekar and Schilling \cite{Ashtekar_etal1998}. Our geometric framework is a natural generalization of general Hopf bundle for pure quantum states, but it is somewhat more complicated than for the pure states. There are some recent works on geometric formulation of the uncertainty relation which are different from our approach which is based on  deep intrinsic geometric structures of quantum phase space of density operators \cite{Hall, Kryukov, Bosyk}.
In section \ref{GfW} we give an short introduction to our geometric framework for mixed quantum states, in section \ref{uncertrel} we derive a geometric uncertainty relation, and in section \ref{ex} we apply the geometric uncertainty relation to a mixture of spin-$\frac{1}{2}$ particles.

\section{Geometry of orbits of isospectral density operators}\label{GfW}
In this paper we consider finite dimensional quantum systems that evolve unitarily.
The systems will be modeled on a Hilbert space $\mathcal{H}$ of unspecified dimension $n$, and their states will be represented by density operators.
Now, the orbits of the left conjugation action of the unitary group $\U(\mathcal{H})$ on the space of density operators on $\mathcal{H}$ are in one-to-one correspondence with the possible spectra for density operators on $\mathcal{H}$, where by the \emph{spectrum} of a density operator of rank $k$ we mean the decreasing sequence
\begin{equation}
\sigma=(p_1,p_2,\dots,p_k)
\label{spectrum}
\end{equation}
of its, not necessarily distinct, positive eigenvalues.
We fix $\sigma$, and write $\mathcal{D}(\sigma)$ for the corresponding orbit of density operators.

To furnish $\mathcal{D}(\sigma)$ with a geometry, let $\L(\mathbf{C}^k,\mathcal{H})$ be the space of linear maps from $\mathbf{C}^k$ to $\mathcal{H}$,
 and $P(\sigma)$ be
the diagonal $k\times k$ matrix
that has $\sigma$ as its diagonal.
Now, we let
\begin{equation}
\mathcal{S}(\sigma)=\{\Psi\in\L(\mathbf{C}^k,\mathcal{H}):\Psi^\dagger \Psi=P(\sigma)\},
\end{equation}
and define
\begin{equation}
\pi:\mathcal{S}(\sigma)\to\mathcal{D}(\sigma),\quad \Psi\mapsto\Psi\Psi^\dagger.
\label{bundle}
\end{equation}
Then $\pi$ is a principal fiber bundle with right acting gauge group
\begin{equation}
\mathcal{U}(\sigma)
=\{U\in\mathcal{U}(k):UP(\sigma)=P(\sigma)U\},
\end{equation}
whose Lie algebra is
\begin{equation}
\mathbf{u}(\sigma)
=\{\xi\in\mathbf{u}(k):\xi P(\sigma)=P(\sigma)\xi\}.
\end{equation}
We  equip $\mathcal{L}(\mathbf{C}^k,\mathcal{H})$ with the Hilbert-Schmidt Hermitian product, and the Riemannian metric $G$ and the symplectic form $\Omega$ given by $2\hbar$ times the real and imaginary parts, respectively, of this product:
\begin{equation}
G(X,Y)=\hbar\mathrm{Tr}(X^\dagger Y+Y^\dagger X), 
\qquad
\Omega(X,Y)=-i\hbar\mathrm{Tr}(X^\dagger Y-Y^\dagger X). \label{Otto}
\end{equation}
We also equip $\mathcal{D}(\sigma)$ with the unique metric $g$ that makes $\pi$ a Riemannian submersion.

The tangent bundle of $\mathcal{S}(\sigma)$ can be decomposed as
\begin{equation}
T\mathcal{S}(\sigma)=\mathrm{V}\mathcal{S}(\sigma)\oplus \mathrm{H}\mathcal{S}(\sigma),
\end{equation}
where
$\mathrm{V}\mathcal{S}(\sigma)=\mathrm{Ker}d \pi$ is the \emph{vertical} and $\mathrm{H}\mathcal{S}(\sigma)=\mathrm{V}\mathcal{S}(\sigma)^\bot$ is \emph{horizontal bundles}
of $T\mathcal{S}(\sigma)$.
Here $^\bot$ denotes orthogonal complement with respect to $G$.
Vectors in $\mathrm{V}\mathcal{S}(\sigma)$ and $\mathrm{H}\mathcal{S}(\sigma)$
are called vertical and horizontal, respectively,
and a curve in $\mathcal{S}(\sigma)$ is called horizontal if its velocity vectors are horizontal.

The infinitesimal generators of the gauge group action yield canonical isomorphisms between $\mathbf{u}(\sigma)$ and the fibers in $\mathrm{V}\mathcal{S}(\sigma)$:
\begin{equation}\label{eq:inf gen}
\mathbf{u}(\sigma)\ni\xi\mapsto \Psi\xi\in\mathrm{V}_\Psi\mathcal{S}(\sigma).
\end{equation}
Furthermore, $\mathrm{H}\mathcal{S}(\sigma)$ is the kernel bundle of the gauge invariant \emph{mechanical connection form}
$\mathcal{A}_{\Psi}=\mathcal{I}_{\Psi}^{-1}J_{\Psi}$,
where $\mathcal{I}_{\Psi}:\mathbf{u}(\sigma)\to \mathbf{u}(\sigma)^*$ and $J_{\Psi}:\mathrm{T}_{\Psi}{\mathcal{S}(\sigma)}\to \mathbf{u}(\sigma)^*$ are the \emph{moment of inertia} and \emph{moment map}, respectively,
\begin{equation}
\mathcal{I}_{\Psi}\xi\cdot \eta=G(\Psi\xi,\Psi\eta),\qquad
J_{\Psi}(X)\cdot\xi=G(X,\Psi\xi).
\end{equation}
The moment of inertia is an adjoint-invariant form on $\mathbf{u}(\sigma)$ which is independent of $\Psi$ in $\mathcal{S}(\sigma)$. Thus
it defines a metric on $\mathbf{u}(\sigma)$:
\begin{equation}\label{eq1}
\xi\cdot \eta=\mathrm{Tr}\left(\left(\xi^\dagger \eta+\eta^\dagger \xi\right)P(\sigma)\right).
\end{equation}
Using equation (\ref{eq1}) we can derive an explicit formula for the connection form.
Indeed, if $m_1, m_2, \dots , m_l$ are the multiplicities of the different eigenvalues in $\sigma$, with $m_1$ being the multiplicity of the greatest eigenvalue, $m_2$ the multiplicity of the second greatest eigenvalue, etc., and if for $j=1,2,\dots,l$,
\begin{equation}
E_j=\mathrm{diag}(0_{m_1},\dots,0_{m_{j-1}},1_{m_j},0_{m_{j+1}},\dots,0_{m_l}),
\end{equation}
then
\begin{equation*}\label{eq:explicit}
\mathcal{A}_\Psi(X)=\sum_jE_j\Psi^\dagger XE_jP(\sigma)^{-1}.
\end{equation*}
Note that the orthogonal projection of $\mathrm{T}_\Psi\mathcal{S}(\sigma)$ onto $\mathrm{V}_\Psi\mathcal{S}(\sigma)$
is given by the connection form followed by the infinitesimal generator given by equation (\ref{eq:inf gen}). Thus the \emph{vertical} and \emph{horizontal projections} of $X$ in $\mathrm{T}_\Psi\mathcal{S}(\sigma)$ are $X^\bot=\Psi\mathcal{A}_\Psi(X)$ and $X^{||}=X-\Psi\mathcal{A}_\Psi(X)$, respectively.

\section{A geometrical uncertainty relation}\label{uncertrel}
The form $\Omega$ given by equation (\ref{Otto}) is a symplectic form on $\L(\mathbf{C}^k,\mathcal{H})$. It follows from a result by Marsden and Weinstein \cite[Th 1]{Marsden_etal1974}, see \cite{Ole2}, that there is a unique symplectic structure $\omega$ on $\mathcal{D}(\sigma)$ such that $\pi^*\omega$ equals the restriction of $\Omega$ to $\S(\sigma)$.
For each observable $\hat A$ on $\mathcal{H}$, define the \emph{expected value function} $A$ and associated \emph{Hamiltonian vector field} $X_A$ on $\mathcal{D}(\sigma)$ by
\begin{equation*}
A(\rho)=\mathrm{Tr}(\hat A\rho),\qquad dA=\iota_{X_A}\omega.
\end{equation*}
Also, let $X_{\hat A}$ be the gauge invariant vector
field on $\mathcal{S}(\sigma)$ defined by
\begin{equation*}
X_{\hat A}(\psi)=\frac{d}{d\varepsilon}\left[\exp\left(\frac{\varepsilon}{i\hbar}\hat A\right)\psi\right]_{\varepsilon=0}.
\end{equation*}
Then $\iota_{\pi_*(X_{\hat A})}\omega=dA$, which means that $X_{\hat A}$ projects onto $X_{A}$.

\begin{figure}[htbp]
\centering
\includegraphics[width=0.6\textwidth,height=0.25\textheight]{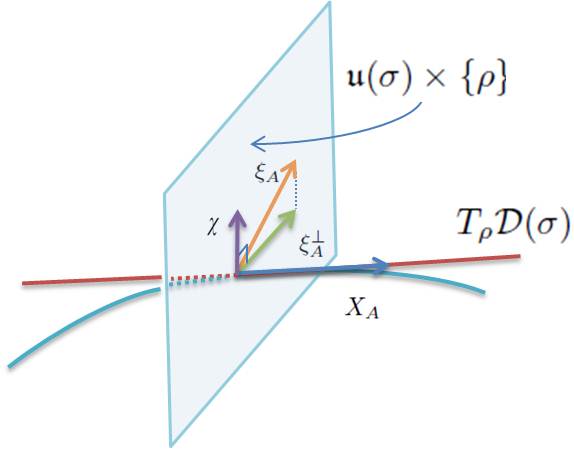}
\caption{Illustration of the $\mathbf{u}(\sigma)$-valued fields $\xi_A$ and $\xi_A^\bot$.}
\label{figure}
\end{figure}

Now, let $\hat{A}$ and $\hat{B}$ be two observables. The \emph{Poisson} and \emph{Riemannian brackets} of their expected value functions are $\{A,B\}_\omega=\omega(X_A,X_B)$ and $\{A,B\}_g=g(X_A,X_B)$. Let $\chi=\1_k/i\sqrt{2\hbar}$. Then
\begin{eqnarray}
&A =\sqrt{\frac{\hbar}{2}}\chi\cdot\xi_A, \qquad B=\sqrt{\frac{\hbar}{2}}\chi\cdot\xi_B,\\
&(A,B) = \frac{\hbar}{2}\left(\{A,B\}_g+\xi_A\cdot\xi_B\right),\qquad [A,B]=\frac{\hbar}{2} \{A,B\}_\omega,
\end{eqnarray}
where $\xi_A$ and $\xi_B$ are the $\mathbf{u}(\sigma)$-valued fields on $\mathcal{D}(\sigma)$ defined by $\pi^*\xi_A=\mathcal{A}\circ X_{\hat{A}}$
and $\pi^*\xi_B=\mathcal{A}\circ X_{\hat{B}}$.
Thus we arrive at the following relation
\begin{equation}\label{covariance}
(A,B)-AB
=\frac{\hbar}{2}\left(\{A,B\}_g+
\xi_A^\bot\cdot\xi_B^\bot\right),
\end{equation}
where $\xi_A^\bot$ and $\xi_B^\bot$ are the projections of $\xi_A$ and $\xi_B$, respectively, on the orthogonal complement of $\chi$, see figure \ref{figure}.

In special case when $B=A$,
\begin{equation}
\Delta A(\rho)^2=(A,A)-AA\geq\frac{\hbar}{2}\{A,A\}_g(\rho).\label{mainett}
\end{equation}
Now, let $X_{\hat{A}}^{||}$ and $X_{\hat{B}}^{||}$ be the horizontal lifts of $X_{\hat{A}}$ and $X_{\hat{B}}$, respectively. Then
the Cauchy-Schwartz inequality applied to the Hilbert-Schmidt Hermitian product gives
\begin{equation*}
G\Big(X_{\hat{A}}^{||},X_{\hat{A}}^{||}\Big)G\Big(X_{\hat{B}}^{||},X_{\hat{B}}^{||}\Big)\geq G\Big(X_{\hat{A}}^{||},X_{\hat{B}}^{||}\Big)^2+\Omega\Big(X_{\hat{A}}^{||},X_{\hat{B}}^{||}\Big)^2.
\end{equation*}
It follows that
\begin{equation}\label{estimat}
\{A,A\}_g\{B,B\}_g\geq \{A,B\}_g^2+\{A,B\}_\omega^2.
\end{equation}
This estimate together with equation (\ref{mainett}) implies
\begin{equation}\label{uncertainty}
\Delta A\Delta B\geq \frac{\hbar}{2}\sqrt{\{A,B\}_g^2+\{A,B\}_\omega^2}.
\end{equation}
We have in detail discussed and compared our geometric uncertainty relation with Robertson-Schr\"{o}dinger uncertainty relation \cite{Ole2}. The advantages of our geometric uncertainty relation for mixed quantum states are the following. Our geometric uncertainty relation is  based on solid and intrinsic geometrical  structures of  underling space of density operators. Moreover, for some class of observables our geometric uncertainty relation perform better than Robertson-Schr\"{o}dinger uncertainty relation. Since our geometric uncertainty relation  depends on $\xi_A^\bot$ and $\xi_B^\bot$ which are intrinsic to the structures of the quantum phase space $\mathcal{D}(\sigma)$. Thus the  application  of our geometric uncertainty relation could give rise to some interesting results e.g., in  the field of quantum information processing. However, for a pure quantum state this geometric uncertainty relation coincides with one derived by Kibble \cite{Kibble_1979}.

\section{Example}\label{ex}
Consider an ensemble of electrons, so prepared that the proportion of electrons with spin up polarization is $p_1$ and the proportion with spin down polarization is $p_2$,
and let $\mathbf{S}$ be the spin-$\frac{1}{2}$ operator. If we model the spin part of the system on $\mathbf{C}^2$ in such a way that $e_1$ and $e_2$ represent the spin up and spin down states, respectively, then the state of the spin part of the ensemble's wave function can be represented by the density operator
$\rho=\left(
       \begin{array}{cc}
        p_1 & 0\\
0 & p_2 \\
       \end{array}
     \right)$
and the components of $\SS$ are:
\begin{equation*}
\hat{S}_x=\frac{\hbar}{2}
\left(
       \begin{array}{cc}
        0 & 1\\
1 & 0 \\
       \end{array}
     \right),\quad
\hat{S}_y=\frac{\hbar}{2}
\left(
       \begin{array}{cc}
        0 & -i\\
i & 0 \\
       \end{array}
     \right),~~
\hat{S}_z=\frac{\hbar}{2}
\left(
       \begin{array}{cc}
        1 & 0\\
0 & -1 \\
       \end{array}
     \right).
\end{equation*}
A lift of $\rho$ to $\S(p_1,p_2)$ is
$
\psi=\left(
       \begin{array}{cc}
        \sqrt{p_1} & 0\\
0 & \sqrt{p_2} \\
       \end{array}
     \right),
$
and the infinitesimal generators of the first two components of $\SS$, evaluated at $\psi$, are
\begin{equation*}
X_{\hat{S}_x}(\psi)=\frac{1}{2i}
\left(
       \begin{array}{cc}
0 & \sqrt{p_2}\\
\sqrt{p_1} & 0\\
       \end{array}
     \right)
,\quad
X_{\hat{S}_y}(\psi)=\frac{1}{2}
\left(
       \begin{array}{cc}
0 & -\sqrt{p_2}\\
\sqrt{p_1} & 0\\
       \end{array}\right).
\end{equation*}
These vectors are horizontal if $p_1\ne p_2$, and vertical if $p_1=p_2$. Regardless, their projections to vectors at $\rho$ are orthogonal. E.g., if $p_1\ne p_2$ we have that
\begin{equation*}
\{S_x,S_y\}_g(\rho)=2\hbar\Re\tr
\left(
       \begin{array}{cc}
ip_1/4 & 0\\
0 & -ip_2/4\\
       \end{array}\right)
=0.
\end{equation*}
Moreover, we have
\begin{equation*}
\{S_x,S_y\}_\omega(\rho)=2\hbar\Im\tr\left(
       \begin{array}{cc}
ip_1/4 & 0\\
0 & -ip_2/4\\
       \end{array}\right)=\frac{\hbar}{2}(p_1-p_2).
\end{equation*}
Consequently,
\begin{equation}
\Delta S_x(\rho)\Delta S_y(\rho)\geq\frac{\hbar^2}{4}(p_1-p_2).
\end{equation}
This example visualize our geometric uncertainty relation in its simplest form. However, it is a straightforward task to determine the relation for arbitrary density operators of rank $k$ defined on a finite dimensional Hilbert state.
\section{Conclusion}
In this paper we have equipped the phase spaces of unitarily evolving quantum systems in mixed states,
with Riemannian and symplectic structures, and we have derived a geometric uncertainty principle for observables acting on quantum systems in mixed states. We  have briefly discussed and compared our geometric uncertainty  relation with other approaches. We have also applied our geometric uncertainty relation to simple physical systems.
Uncertainty relations have found many application in the field of quantum information processing.
The rich geometric structure of  our uncertainty relation indicates that it could have many applications in quantum information. However, this issues  needs further investigations.
\begin{flushleft}
\textbf{Acknowledgments:}
 This work was supported by the Swedish Research Council (VR).
\end{flushleft}


\end{document}